# The spectral evolution and ejecta of recurrent nova U Sco in the 2010 outburst

M. P. Diaz[1], R. E. Williams[2], G. J. Luna[3], M. Moraes[1] and L. Takeda[1]

[1] IAG, Universidade de São Paulo
Rua do Matão, 1226
São Paulo, SP, 05508-900, Brazil
marcos@astro.iag.usp.br

[2] Space Telescope Science Institute
3700 San Martin Drive
Baltimore, MD 21218, USA

[3]Center for Astrophysics / Smithsonian Astrophysical Observatory
60 Garden st. MS15,
Cambridge, MA 02138, USA

ABSTRACT

Synoptic spectroscopic observations of the U Sco 2010 outburst from maximum light to quiescence as well as a contemporaneous X-ray observation are presented and analyzed. The X-ray spectrum 52 days after outburst indicates a hot source ($kT_{bb} \sim 70$ eV). Narrow line components from the irradiated companion atmosphere were observed in hydrogen and helium optical recombination lines. The formation of a nebular spectrum is seen for the first time in this class of recurrent novae, allowing a detailed study of the ejecta using photoionization models. Unusual [OIII] auroral-to-nebular line ratios were found and possible scenarios of their origin are discussed. The modeling of the emission line spectrum suggests a highly heterogeneous ejecta with mass around or above $3 \times 10^{-6}$ $M_{sun}$.

1. INTRODUCTION

Recurrent novae are a subclass of eruptive cataclysmic variables (CVs) for which more than one nova outburst has been recorded (see Warner 1995 for a review). The repeated outbursts in those objects offer the possibility of better understanding the nova phenomenon by comparing pre- and post-outburst properties and sampling differences between outbursts. The recurrent nova activity has been interpreted as a possible signature of the final evolution of a CV towards a type Ia supernovae, with the short recurrence times scales caused mostly by a high mass white dwarf. In this scenario, the fate of the system is ultimately defined by the secular balance between mass accretion in quiescence vs. mass loss in recurrent nova outbursts. These novae are possibly the most promising SN Ia progenitor candidates among non steady-burning white dwarfs (Di Stefano 2010). Uncertainties in mass transfer rates, mass loss in accretion disk winds and ejected masses in nova outbursts all add to the uncertain evolution of recurrent novae.

The long period, eclipsing (Schaefer & Ringwald 1995) recurrent nova U Sco exhibits fast, large photometric amplitude outbursts (Schaefer 2010) associated with extremely high shell expansion velocities (>4000 km.s$^{-1}$). It has been considered the prototype of the SNIa progenitors among normal CVs. These systems harbor a white dwarf whose mass is close to the Chandrashekar limit and a slightly evolved, less massive companion (Thoroughgood et al 2001, Hachisu et al. 2000a,b). The envelope mass in the 1999 outburst of U Sco was determined to be 3x10$^{-6}$ M$_{sun}$ from its light curve modeling (Hachisu et al. 2000a) while the shell mass was estimated as 10$^{-7}$ M$_{sun}$ from recombination line fluxes (Anupama & Dewangan 2000) and 2.7(0.9) x 10$^{-7}$ M$_{sun}$ from the IR free-free continuum (Evans et al. 2001). Previous (1979) outburst ejected mass are also in the range of $10^{-6}$ to $10^{-7}$ $M_{sun}$ (Williams et al. 1981). On the other hand, mass accretion rates estimates are more heavily dependent on models, yielding figures as high as $2.5 \times 10^{-7}$ $M_{sun}$ $yr^{-1}$ (Hachisu et al. 2000a) with significant steady mass loss from the system as inferred from the P-dot value found by Matsumoto, Kato & Hachisu (2003).

Analysis of the ejected mass indicate depletion of hydrogen in a helium rich shell with He/H abundance ratios estimated to range from 2 to 4 (Williams et al 1981, Evans et al. 2001). This finding is consistent with the rapid cooling and optical decline (Hachisu et al. 2000a). However, contradicting values of 0.16 and 0.4 were also found for the 1999 ejecta using optical lines (Iijima 2002; Anupama & Dewangan 2000). Nitrogen and oxygen were also found to be overabundant in the ejecta, as expected from fast novae thermonuclear runaway (TNR) simulations (Starrfield et al. 2009).

Early and brief supersoft X-ray emission was observed in previous outburst of this recurrent nova during 1999 (BeppoSAX) and 2010 (SWIFT), with onset as soon as 20 days after maximum, i.e., in the midplateau phase, and lasting typically 20 to 30 days (Kahabka et al. 1999, Schlegel et al. 2010). While estimated X-ray luminosities range from $10^{36}$ to $10^{37}$ $erg.s^{-1}$ during the nuclear burning phase, no nebular spectrum has been observed in previous outbursts of U Sco-type recurrent novae (Warner 1995).

## 2. OBSERVATIONS

### 2.1 *Optical Spectrophotometry*

The optical observations of U Sco presented here were taken as part of the SOAR Telescope Synoptic Survey of Novae, which allows for Target of Opportunity spectrophotometric monitoring of confirmed novae from outburst maximum down to quiescence. The combination of a 4.2 meter telescope with the capabilities of the GOODMAN high throughput spectrograph (Clemens, Crain and Anderson, 2004), yield sufficient S/N spectra with resolution R~2800. Two consecutive exposures using a 600 l/mm VPH grating with different wavelength settings produce an extended optical coverage from 3500 Å to 9000 Å with a small gap around 6300 Å. The instrumental profile is well sampled by the 15x15 microns pixels of the Fairchild 4k x 4k UV-optimized CCD. Individual exposure times from 3 to 25 minutes were used. A narrow slit of 0.46 arc seconds was employed to achieve the maximum spectrograph resolution with full optical coverage in two exposures. Tertiary standard calibration stars from the list of Hamuy et al. (1994) were observed employing wide slits widths under photometric conditions on each visit. Target exposures were followed by Hg-Ar frames. Projected flats in the red were taken during the night for better fringing correction, however, substantial residual fringing persisted redward of the telluric A band. The data reduction using IRAF[1] packages included usual bias, flatfielding procedures and standard extraction of the spectra, followed by wavelength and flux calibration. A total of 26 spectra of the science target, including the wide slit spectra used for spectrophotometric calibration, were obtained over a period of six months.

---

[1] IRAF is distributed by the National Optical Astronomy Observatory, which is operated by the Association of Universities for Research in Astronomy, Inc., under cooperative agreement with the National Science Foundation.

2.2 *X-ray observations*

The SWIFT satellite monitored the latest U Sco outburst, approximately every day beginning January 29, 2010. The X-ray observation closest to the date of one of our optical observations was taken on March 21, 2010 (ObsID: 00031417033) using the XRT (X-Ray Telescope) in PC (Photon Counting) mode for 5.4 ks. We extracted the spectrum of the source within a circular region of 20 pixels radius, and background from an annulus region of 26 and 40 pixels inner and outer radius, respectively, using the HEASOFT[1] tool XSELECT. We created ancillary matrices (ARF) using the tool XRTMKARF, and used the response matrices (RMF) available through the CALDB[2] database. Finally, the source spectrum was binned at a 5 counts per bin minimum.

3. THE OPTICAL AND X-RAY SPECTRA

3.1 S*pectral evolution*

The rapid decay of U Sco was sampled by 7 visits beginning less than one day after its discovery on February 28.4 ($t_0$) until the nova had decayed to its quiescent magnitude (V ~ 18.8) in July 2010 (fig.

---
1  http://heasarc.gsfc.nasa.gov/docs/software/lheasoft/
2  http://heasarc.gsfc.nasa.gov/docs/heasarc/caldb/data/swift/xrt/

1). Changes in the optical spectrum occurred rather quickly in the early decline period, with significant modifications in the line profiles, ionization and continuum shape taking place over time-scales of a few days. A steep Balmer decrement and HeI lines are present in the spectrum near maximum light. After maximum the spectrum evolved from a $P_N^O$ class (Williams, 1992) on January 29.3 and 30.3 to the flat continuum $P_N$ spectrum seen on February 6, with the appearance of narrow Balmer and HeII (fig. 2) components and significantly weaker HeI emission. This suggests an increase in the shell ionization during this period. The continuum was red in the first spectrum changing to flat by January 30, and becoming blue by February 6. Subsequently, the broad permitted lines weaken and an almost pure Balmer and HeII narrow line spectrum ($P_{He}$) superposed on a bright UV continuum is present by February 20 ($t-t_0 = 23$ d). Weak high ionization lines of NV and OVI where tentatively identified at this time (fig. 2).

In a departure from its behavior in previous outbursts, the spectrum evolved to a nebular type ($N_O$) between February 20 ($t-t_0 = 23$ d) and March 20 ($t-t_0 = 51$ d). Broad oxygen, nitrogen and neon forbidden lines are seen (fig. 3) over a much flatter blue continuum which may possibly be due to renewed accretion disk emission. A weakening of the narrow Balmer + HeII lines continues during this period. Extremely high auroral-to-nebular line intensity ratios are displayed by both [OIII] and [NII] lines, indicative of collisional depopulation of $^1D_2$ upper levels in a high density environment. These spectral features were confirmed by a later observation on April 13 ($t-t_0 = 75$ d), when a decrease in the 4363/5007 ratio was noticed in a fainter $N_O$-type spectrum, which is consistent with a decrease in electron density due to shell expansion. He I lines continue to be absent in the late decline period. The [NeIII] lines in the UV are observed to be bright in both nebular spectra. This represents the first identification of these lines in any U Sco spectra.

The last spectra of U Sco taken in our observing campaign on July 10 ($t-t_0 = 163$ d) occurred at B ~

19.1 and showed weak HeII lines, resembling those of the quiescent spectrum, as well as residual broad nebular lines (fig. 3). The [OIII] 4363 Å and [NII] 5755 Å lines had disappeared, suggesting a lower density and possibly cooler gas. The observed line fluxes during the nebular phase are given in table 1.

Narrow NaI D and CaII H+K absorption lines could be observed during postoutburst. On January 30 the interstellar/circumbinary NaD had measured equivalent widths of EW(D1)=320(30) mÅ and EW(D2)=190(30) mÅ. Similar values were found in the other spectra. Measurements of H+K features yield variable equivalent widths possibly due to blending with nearby varying emission lines.

### 3.2 *Line profiles and the narrow component spectrum*

The broad lines seen close to maximum show structured saturated profiles with FWHM ~7500 km.s$^{-1}$ and FWZI ~9200 km.s$^{-1}$. A symmetric absorption system is also seen in those lines at +/- 2400 km.s$^{-1}$, suggesting a velocity gradient in the ejecta. The OI 8446 fluorescence excited line, which roughly traces the neutral gas velocity distribution in the ejecta, presents a different U-shaped profile with comparable widths. The typical post-maximum profiles turn into composite profiles in less 5 days, displaying a narrow (500 to 600 km.s$^{-1}$) component for hydrogen, helium and possibly the NIII 4640 lines. Contrary to the spectrum during initial decay, the narrow component shows strong helium and weak hydrogen emission with flat Balmer decrements. Possible explanations for the origin of such narrow lines are (1) the reionization of circumbinary gas from previous outbursts, or (2) chromospheric emission from x-ray illumination of the companion by the shrinking nova photosphere. The resolved narrow component profiles are shown in detail in figure 4. All wavelength calibrations were rectified using the Na D absorption lines as well as the telluric $O_2$ B-band head and auroral lines. Measured velocities of narrow components have been determined from their flux weighted centroids as measured on February 6. Significant radial velocity variations are seen, with amplitudes consistent with the orbital velocity $K_2$ of this high-inclination system (Thoroughgood et al 2001). This is strongly indicative of an origin associated with the companion star.

### 3.3 *Analysis of the X-ray emission*

The X-ray spectrum of U Sco from March 21 extends from 0.2 to 2.0 keV, however the source emission from ~1.0 to 2.0 keV is only ~2 sigma above the background level. With this in mind, we fit the X-ray spectrum with two different models. First, we fit the 0.2-2.0 keV spectrum using an absorbed black-body plus a thin thermal plasma model using XPEC[1] (with $\chi^2_v$ = 0.80 and 12 d.o.f.), representing the emission from the nova photosphere and likely shocked thermal emission (figure 5a). This fit yields a temperature for the black-body emission of $kT_{bb}$ = 65±30 eV; an unabsorbed luminosity of $L_X$ = $1.4 \times 10^{35}$ $(d/10 \text{ kpc})^2$ erg.s$^{-1}$, where *d* is the distance to U Sco; an absorber column density $n_H$ = 2.0(+0.5)(-2.0) x $10^{21}$ cm$^{-2}$; a temperature for the thin thermal plasma $kT_{th}$ = 350(170) eV and normalization for the thermal plasma of 5.5(1.6) x $10^5$. For the second model, we used an absorbed pure black-body in the 0.2-1.0 keV energy range with $\chi^2_v$ = 0.71 and 11 d.o.f, resulting in the following parameters: black-body temperature $kT_{bb}$ = 72(+40)(-20) eV; unabsorbed luminosity $\log(L_X)$ = 35.0(+2.1)(-1.1) $(d/10 \text{ kpc})^2$ erg.s$^{-1}$ and absorption column density $n_H$ = 2.3 -- 4.3 x $10^{21}$ cm$^{-2}$. (figure 5b). The two models applied yield comparable black-body luminosities and temperatures. However, the uncertainties in the effective ionizing fluxes in the soft X-ray range are high, mainly due to the extrapolation of the observed X-ray spectrum towards lower energies with the assumption of a black-body SED. This is the reason for leaving the temperature and luminosity of a black-body equivalent ionizing source as free parameters in the photoionization model grids.

---

[1] http://heasarc.gsfc.nasa.gov/docs/xanadu/xspec/

# 4 PHOTOIONIZATION MODELS

## 4.1 *Model parameters*

The onset of a nebular phase for this nova gives us additional diagnostic capabilities by using photoionization simulations. The spectra taken on March 20 and April 13 were used to determine line fluxes. On those dates, [OIII], [NII] and [NeIII] nebular and auroral lines are present, as well as HeII and Balmer recombination lines. However, severe blends involving Hβ and Hγ prevented us from using those lines. Hα is also blended with [NII] although, as explained below, its contribution to the line flux is expected to be small. Upper limits for bright forbidden transitions where also used to constrain the models. Contributions from a reestablished accretion disk to the recombination lines may also be present by April 13. However, the observed line FWHM and shape are inconsistent with those seen in the system during quiescence.

The distance to the system, which is essential to determine the emission measure, is uncertain with published values of 4(2), 7.5, 12(2) kpc (Hanes 1985, Hachisu et al. 2000a,b, Schaefer 2010. see also the discussion in Webbink et al. 1987). The later value was derived using an evolved subgiant companion spectroscopic parallax during the eclipses and it may be more accurate, requiring a knowledge of the evolutionary state of the secondary star. The above distances place U Sco in the galactic halo. Reddening by Galactic and circumbinary gas has been estimated as E(B-V) = 0.13(3) using the empirical relation between EW(NaD) and E(B-V) of Munari and Zwitter (1997). Such a value is consistent with the $N_H$ range of 1.8 -- 4.8 x$10^{21}$ cm$^{-2}$ estimated by Kahabka et al. (1999), and within the values predicted by galactic extinction models; E(B-V) = 0.09, 0.36 and 0.24 by Amores and Lepine (2005), Schlegel, Finkbeiner and Davis (1998) and Burstein and Heiles (1982), respectively. A correction for E(B-V) = 0.15 was adopted for both lines and continuum fluxes.

A lower limit to He/H abundance can be calculated from the intensities of Hα and HeII λλ 4686, 5411, 4199, 4541 Å for both March 20 and April 13, and the resulting values of He/H by number range from 0.1 to 0.2. However, uncertainties are introduced by the assumption of case-B recombination in a dense medium (see next section), in addition to the already mentioned uncertainties in the Hα flux. Therefore, a wide range of helium abundances is possible.

An extensive grid of shell simulations was performed attempting to reproduce the observed line fluxes, for which a wide range of input parameters was assumed, namely: the ionizing source luminosity $\log(L_*)$ = [35-38.5] erg.s$^{-1}$ and temperature $T_{bb}$ = [3-8x$10^5$] K, the shell mass $M_{shell}$ = [$10^{-7}$- 2x$10^{-5}$] $M_{sun}$ and helium abundance He/H = [0.1- 2]. Average CNO nova abundances from the compilation by Gerhz et al. (1998) were used for most models. The distance was allowed to take on values between 4 -- 13.5 kpc, being scaled to best match the observed line fluxes. The lower limit of 4 kpc was chosen to account for a possible overestimate of the Hα due to its blend with [NII]. Given the uncertainty in the spectral energy distribution of the central source below the Lyman continuum (which mostly defines the model behavior), a simple black-body ionizing continuum was employed for all models. Spherical models were computed with the shell radius and thickness given by the average HWZI of emission lines and the slower absorption system velocities found in the previous section, combined with the appropriate time after outburst and assuming free expansion.

The calculations were performed using RAINY3D (Diaz, 2002) which drives the photoionization code CLOUDY 06.02b (Ferland et al. 1998) to produce 3D models of optically thin shells with condensations. Both homogeneous and heterogeneous models were computed. In clumpy models a pseudo-3D shell is assembled from a 1D mesh. The 3D shell is described by a spherical grid with adaptive solid angle elements, with 1D models being computed along each direction from the central source. The basic calculations, including the photoionization and thermal equilibrium, in those clumpy models are therefore performed inside the 1D code CLOUDY. The resulting line emissivities are then integrated over the whole shell or in selected regions. The transfer of diffuse radiation inside the nebula is made along the radial direction or assuming on-the-spot approximation.

### 4.2 Clumpy shell models

The presence of discrete clumps in the novae ejecta is observed in spatially resolved shells, e.g., HR Del, GK Per, RR Pic and T Pyx. For novae without resolved shells, certain forbidden emission line ratios can be explained only if the ejecta have a very inhomogeneous mass distribution (Williams, 1992). There are several mechanisms that can give rise to condensations in the expanding ejecta. Rayleigh-Taylor instabilities (RTI) must certainly be excited in the very early expansion. Subsequently, as the shell density decreases, becoming optically thin to emission lines, Kelvin-Helmholtz instabilities should also develop in the gas (Chevalier et al 1992). The onset of thermal instabilities depends on the properties of the nova, e.g., chemical composition, matter distribution, central source luminosity and temperature and can also form clumps in the shell (Pistinner and Shaviv, 1995). The binary motion and white dwarf rotation were included in 2.5D hydrodynamical calculations by Lloyd et al (1997), who found a density contrast of 64 between the RTI-formed globules and the neighborhood diffuse regions. A density contrast of 3 to 16 was observed in the evolved HR Del ejecta (Moraes & Diaz, 2009). Lloyd et al. (1997) also found a relation between the maximum condensation size and light curve decay speed ($t_3$) for RTI formed condensations. For very fast novae, $t_3 < 12$ days, the maximum clumping length is ~0.4 times the shell radius, e.g., V1500 Cyg, (Slavin et al. 1995). U Sco has $t_3 = 2.6$ days and therefore its shell may harbor a number of large globules. Taking into account its fast evolution, the presence of numerous small clumps caused by thermal instability may be considered as well. Since the U Sco shell has a high expansion velocity and the orbital period is long, the binary motion should not affect the mass distribution of the ejecta (Livio et al. 1990).

In the case of an unresolved shell like U Sco the condensation properties are cannot be determined from observations. Shell models with a random distribution of Gaussian gas globules with sizes (2xFWHM) between 10% and 80% of the shell radius were computed for 30%, 70% and 80% of the total mass contained in those condensations. Density contrasts up to 100 were tried as an additional input parameter to the simulations. Those 3D models require long processing times for completion and the parameter space searched had to be restricted.

### 4.3 *Model results*

For U Sco few line fluxes and useful upper limits are available to constrain the models. Because of the small number of degrees of freedom due to the large number of poorly constrained parameters our simulations were aimed only at obtaining limits to some of the shell physical parameters, enable us to reject model families and identify plausible scenarios for the production of the nebular spectrum. Trying to constrain the most important parameters, given the model uncertainty in the ionization structure due to unknown abundances and mass distribution, we have simply plotted all models matching the total observed line flux in the ionizing luminosity versus shell mass plane (figure 6). Alternatively, all models matching the Hα luminosity range are shown to the same scale. The computed model grids suggest that the mass ejected in this outburst is larger than ~$8 \times 10^{-7}$ $M_{sun}$, and also that significant Balmer self-absorption occurs during the nebular phase thus, departing from case B recombination.

The high [OIII] auroral-to-nebular ratios could not be reproduced by any homogeneous model with shell mass below $10^{-5}$ $M_{sun}$, even when the shell thickness is arbitrarily reduced to of 1/8 of its radius, and with density falling as $(t-t_0)^{-2}$. This is due to the fact that the high expansion velocities yield a low density gas, while near critical densities are required to produce the observed line ratios. Models with condensations can explain the observed oxygen line ratio at lower shell masses. For instance, a clumpy model with few large condensations can reproduce the [OIII] line fluxes and the auroral-to-nebular ratio seen on April 13 with a total shell mass of $3 \times 10^{-6}$ $M_{sun}$, a central source with $T_{bb}$ = 800,000 K, and $\log(L_*)$ = 37.5 erg.s$^{-1}$. However, ionization of clumpy shells by higher luminosities and lower temperatures yields recombination line fluxes well above the observed values due to the quadratic density dependence appropriate for high ionization parameters. The line ratios derived from the simulations also indicate that black-body temperatures $T_{bb}$ < 500,000 K and/or luminosities below $10^{35.5}$ erg.s$^{-1}$ are unlikely to occur during the onset of the nebular phase.

The critical densities for the [NII] nebular trasitions are more than 10 times lower than the corresponding [OIII] lines (Osterbrock and Ferland, 2006). Therefore, considering that [OIII] 5007,4959 Å lines are already attenuated by collisions one may expect faint [NII] nebular emission. For typical high density conditions in our models the flux ratio 6548+6584 / 5755 is less than ~0.3. The [NeIII] 3869 Å line seen in the nebular phase is also possibly dimmed by collisional effects. If both [OIII] and [NeIII] (with 35 eV and 41 eV ionization potentials, respectively) lines are formed in the same regions inside the nebula, then a significant neon overabundance, greater than ~1 dex, relative to the solar value is needed to explain the observed line flux.

## 5. DISCUSSION

The post-maximum V light curve of the 2010 outburst (fig. 1) resembles that observed in 1987, but it deviates from other outburst curves (Schaefer, 2010). In particular, the midplateau phase is less step in the 2010 outburst, presenting a well defined start at a fainter magnitude. In the July spectra the system had returned to its normal quiescent magnitude and, except for residual nebular emission, to its featureless pre-outburst spectrum as well.

Regardless of the similarity in postoutburst light curves, the current spectral evolution of U Sco displayed certain differences from previous outbursts. For instance, the 2010 outburst displayed recombination line profiles that rapidly became sharp and single peaked less than 9 days after maximum, in contrast with the evolution in 1999, where they remained highly structured at least 17 days after maximum (Iijima 2002). It is also interesting that the 1979 early outburst spectra (Mallama & Starosta 1980, Barlow et al. 1981) show much lower Balmer to helium line ratios when compared to the 2010 spectra taken at $(t-t_0) = 1$ and 2 days, in spite of the great similarity of the 2010 spectrum at maximum light when compared with the 1999 outburst (Kafka and Williams 2010). These observations taken together with the parameters constrained by our models suggest that the central source and shell properties are distributed over a significant range among recurrent nova episodes. The origin of differences between outbursts and also between recurrence intervals is unknown. In the case of U Sco one may speculate that they are related to mass transfer variations induced by the long-term stellar variability of the evolved companion.

High ionization lines of CIV, NV and OVI have previously been identified in the spectrum of U Sco in outburst (Barlow et a. 1981). Some of those lines are also tentatively identified in 2010, during a phase when lower ionization metal lines are weak or absent. This indicates a period of high and complete ionization of the ejecta. A nebular phase could be identified and followed for the first time in this type of recurrent novae. This may be due to an extended X-ray bright phase in this outburst, lasting until the shell has reached nebular densities. Alternatively, one has to consider the strong selection effect is imposed by the sparse spectroscopic sampling beyond the 60 th day in previous outbursts. A late, brief (~3 months) and faint nebular phase as seen in this outburst would have been difficult to identify in prior outbursts.

The photoionization simulations presented here suggest that this outburst was also different from 1979 and 1999 outbursts in terms of the mass loss from the system. While previous estimates point toward shell masses around $10^{-7}$ $M_{sun}$, an ejected mass above $\sim 3 \times 10^{-6}$ is derived from our analysis of the 2010 shell. A long term average mass accretion rate of less than $2.5 \times 10^{-7}$ $M_{sun}.yr^{-1}$ was estimated by Matsumoto, Kato and Hachisu (2003). On the other hand, the U Sco recurrences have been quite regular. With the assumption of missed outbursts in ~1926 and ~1955 their average spacing is 10.2 years. By assuming constancy of current parameters, one finds that the evolution of the system towards a SNIa is doubtful if ejections like that experienced by the white dwarf in 2010 are frequent among the nova recurrences. On the other hand, the observation of bright neon lines may suggest that the primary is a high mass ONeMg white dwarf that remained below the Chandrasehkar during its accretion history. Better data in the UV and more precise modeling of the shell would be needed to confirm an overabundance of neon comparable to those found in well known neon novae.

## 6. CONCLUSIONS

Synoptic optical observations of U Sco during its 2010 outburst show a peculiar spectral evolution with the development of an oxygen-type nebular phase. Almost simultaneous X-ray observations at day 52 suggest a hot central ionizing source with $T_{bb}$ = 840,000 K or slightly lower if a thermal plasma is included in the fit. Both the observed spectra and the photoionization models suggest that the 2010 U Sco outburst may be different from previous outbursts in some aspects. Extreme auroral-to-nebular line intensity ratios were observed during the nebular phase, helping to constrain the shell models to relatively high masses. The shell mass is estimated to be above $3 \times 10^{-6}$ $M_{sun.}$, probably distributed in highly structured, inhomogeneous ejecta. Transient [NeIII] forbidden lines were observed suggesting an overabundance of neon that could exceed ~1 dex above the solar value. Narrow recombination line components detected during the decay towards minimum light are interpreted as the chromospheric emission from the irradiated companion.

We thank the SOAR resident astronomers Tina Armond, Sergio Scarano and Luciano Fraga for the careful queue mode observations. MPD acknowledges the support by CNPq under grant #305725. We acknowledge with thanks the variable star observations from the AAVSO International Database contributed by observers worldwide and used in this research. This research was based on data obtained at the SOAR telescope. This research has made use of data obtained from the High Energy Astrophysics Science Archive Research Center (HEASARC), provided by NASA's Goddard Space Flight Center.

FIGURE CAPTIONS

Figure 1. Light curve of U Sco in the V band according to AAVSO photometric observations. The pluses and triangles represent HeII 4686 Å and Hα+[NII] fluxes on an arbitrary magnitude scale. The arrows show the spectroscopic sampling of the nova decay. The Swift/XRT observations analyzed in this work were taken at $(t-t_0) = 52$ days.

Figure 2. Extended optical spectra of U Sco. The time from maximum is indicated as well as suggested line identification. The spectral resolution range from 1.3 to 2.8 Å (FWHM). Intensity unit is $f_\lambda$ (erg.cm$^{-2}$.s$^{-1}$. Å).

Figure 3. Same as figure 2 for the nebular phase data. The resolution of the July 10 spectrum was degraded by Gaussian convolution to R~1400 for plotting.

Figure 4. Narrow component of recombination lines in U Sco. The broad profile has been subtracted and the all intensities have been normalized to have the same maximum flux. The dotted lines represent the flux weighted centroid of Hα and HeII on February 6 spectra.

Figure 5. Swift/XRT spectra from U Sco obtained on March 21 $(t-t_0 = 52$ days). Panel (a) shows the spectrum modeled with an absorbed black-body plus thermal component (dotted lines) while panel (b) shows the spectrum modeled only with an absorbed black-body. The contribution of the thermal component is only evident in the higher energies where the measured flux is only ~2σ above the background, nevertheless, the luminosity of the black-body component remains compatible between the two models (see text for details). Fit residuals are shown in the bottom panels.

Figure 6. Slice of the multidimensional parameter space of shell photoionization models for $(t-t_0) = 51$ days. The open squares represent clumpy models while diamonds are homogeneous spherical shells. The model emission line emissivity is integrated over the shell to obtain the line luminosity. The top panel shows all models that reproduce the total optical emission line luminosity (i.e. all observed lines summed) while the bottom panel presents all models that satisfy the Hα luminosity, for a wide distance range (see text). Note the lack of solutions on the left side of each panel.

| TABLE 1 | | | |
|---|---|---|---|
| Observed nebular phase line fluxes[a] | | | |
| Line | (t-$t_0$) = 51 d | (t-$t_0$) = 75 d | (t-$t_0$) = 163 d |
| Hα + [NII] 6548/84 Å | 446 | 187 | 110 |
| [NII] 5755 Å | 172 | 71 | 15 |
| HeII 5412 Å | 11.2 | 3.4 | - |
| [OIII] 5007/4959 Å | 1040 | 526 | 48 |
| HeII 4686 Å[b] | 220 | 49 | 7[c] |
| HeII 4542 Å | 5.0 | 1.8 | - |
| [OIII] 4363 Å + Hγ | 400 | 110 | <20 |
| HeII 4199 Å | 3.1 | 2.2 | - |
| Hδ | 60 | 7.5 | - |
| [NeIII] 3869 Å | 110 | 44 | - |
| CaII H | - | - | 2.2(4)[d] |
| CaII K | - | - | 2.4(4)[d] |

[a] Flux uncertainties range from 10% to 30%. Units are $10^{-16}$ erg.cm$^{-2}$.s$^{-1}$.
[b] Broad component.
[c] Narrow component.
[d] absorption equivalent width (Å).

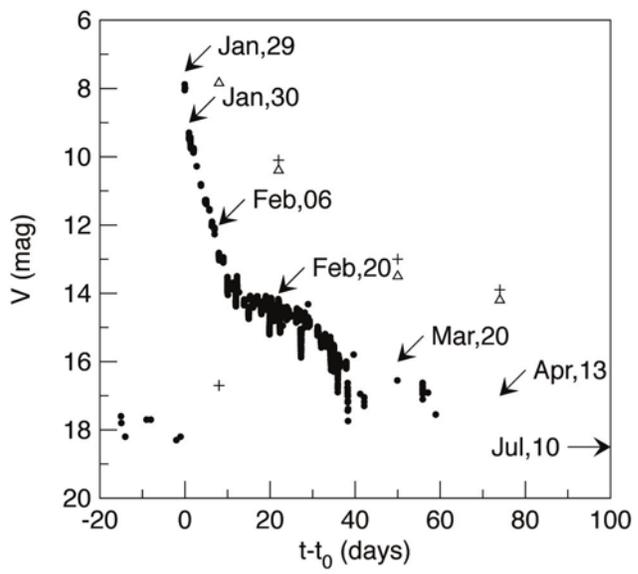

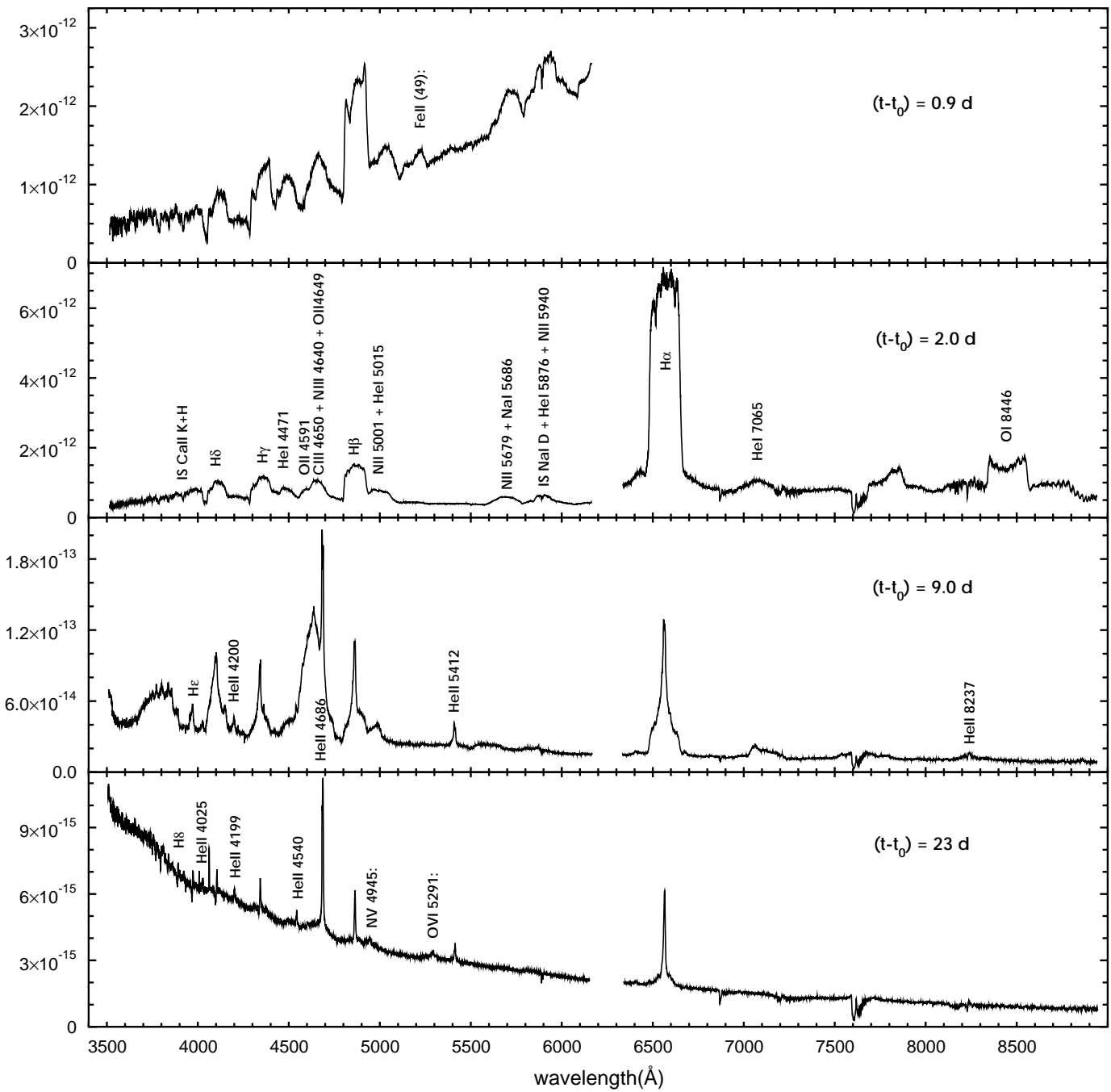

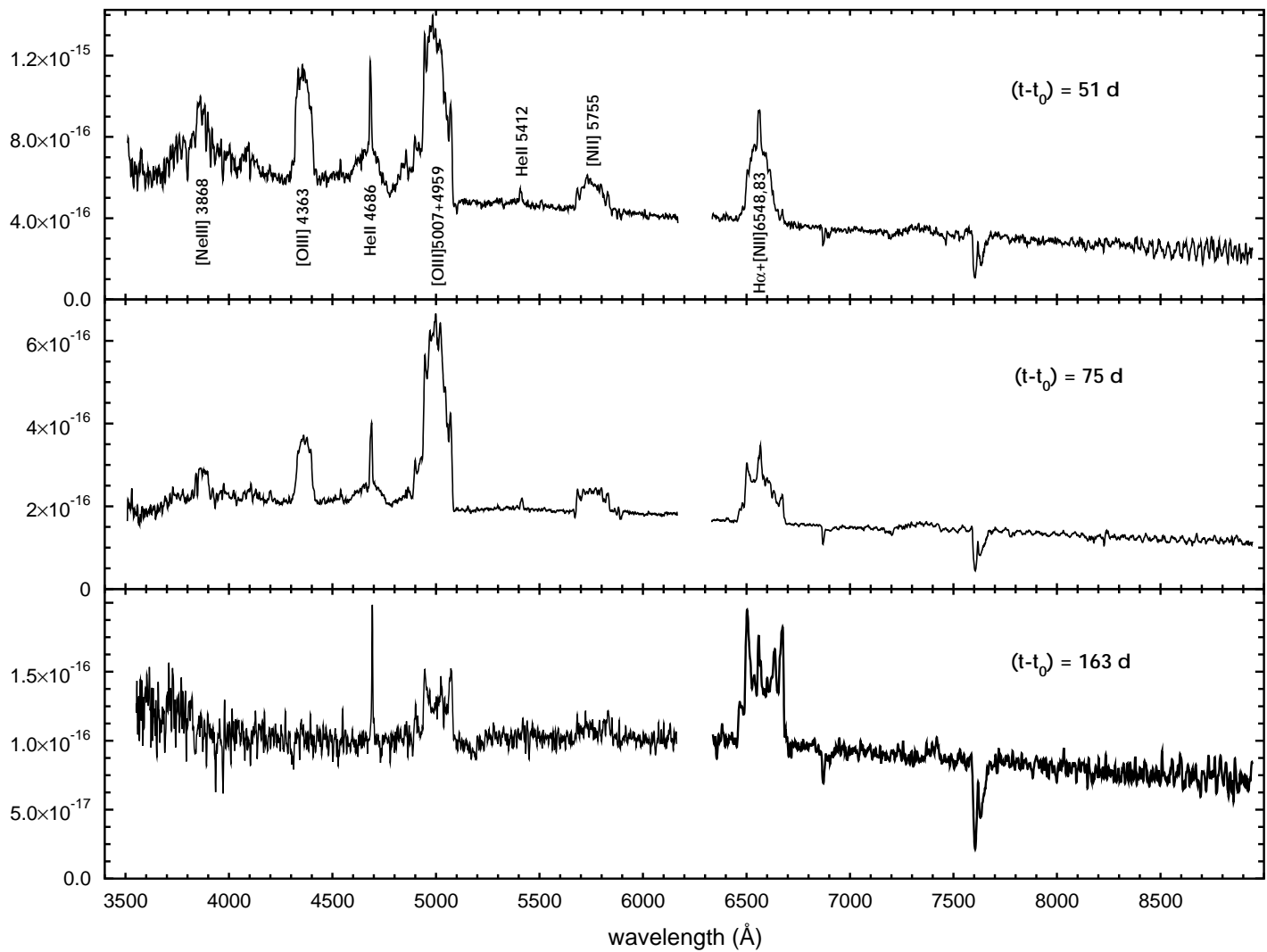

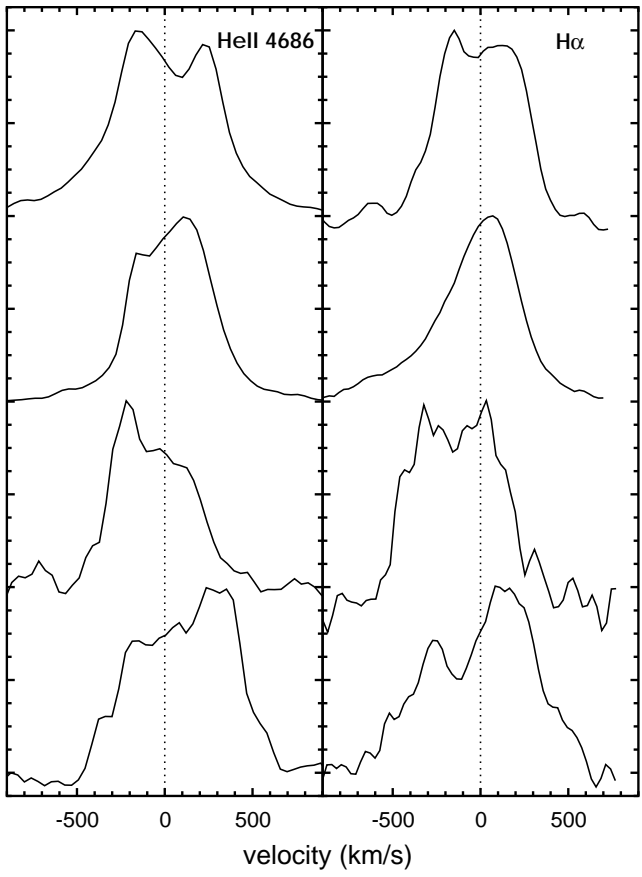

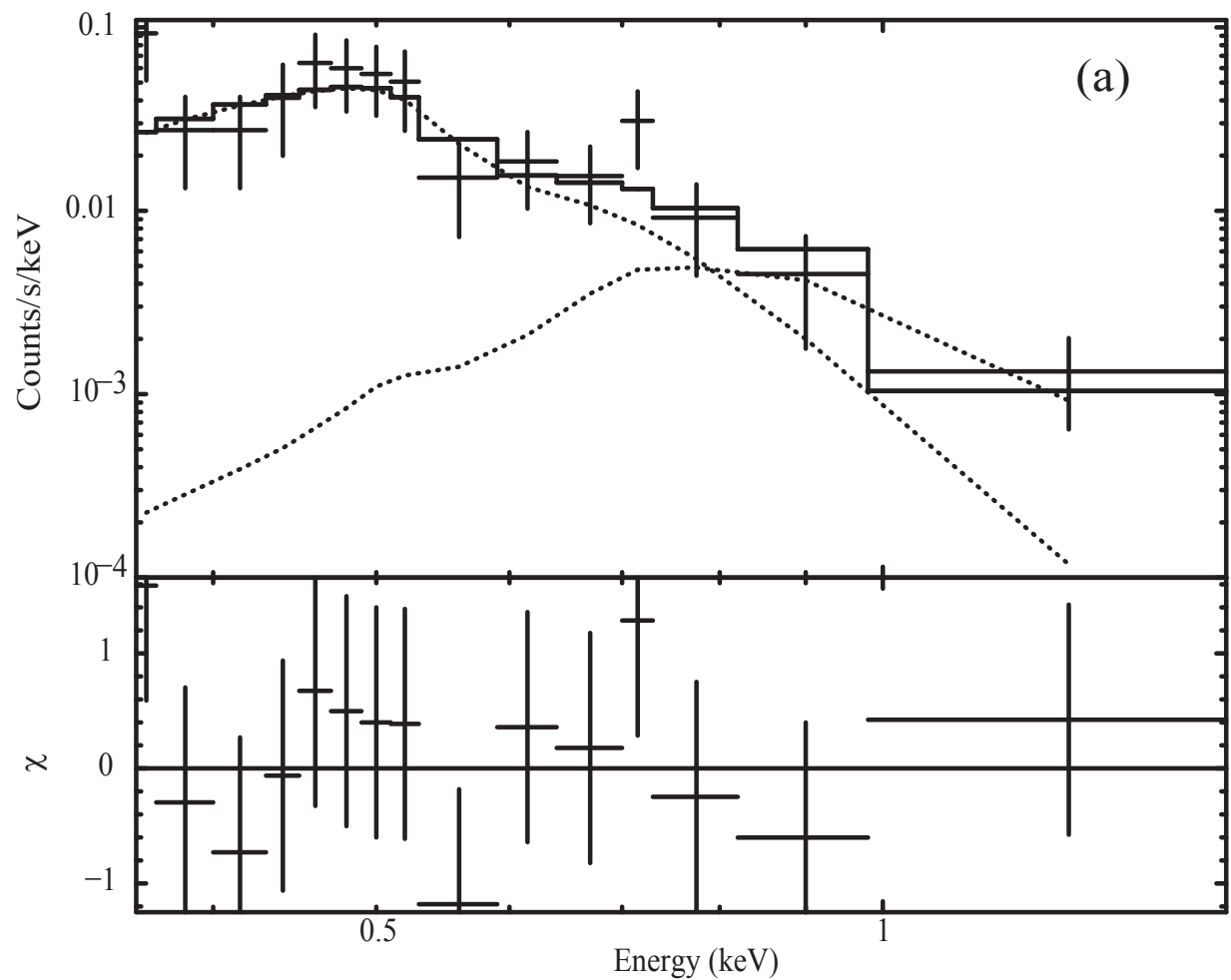 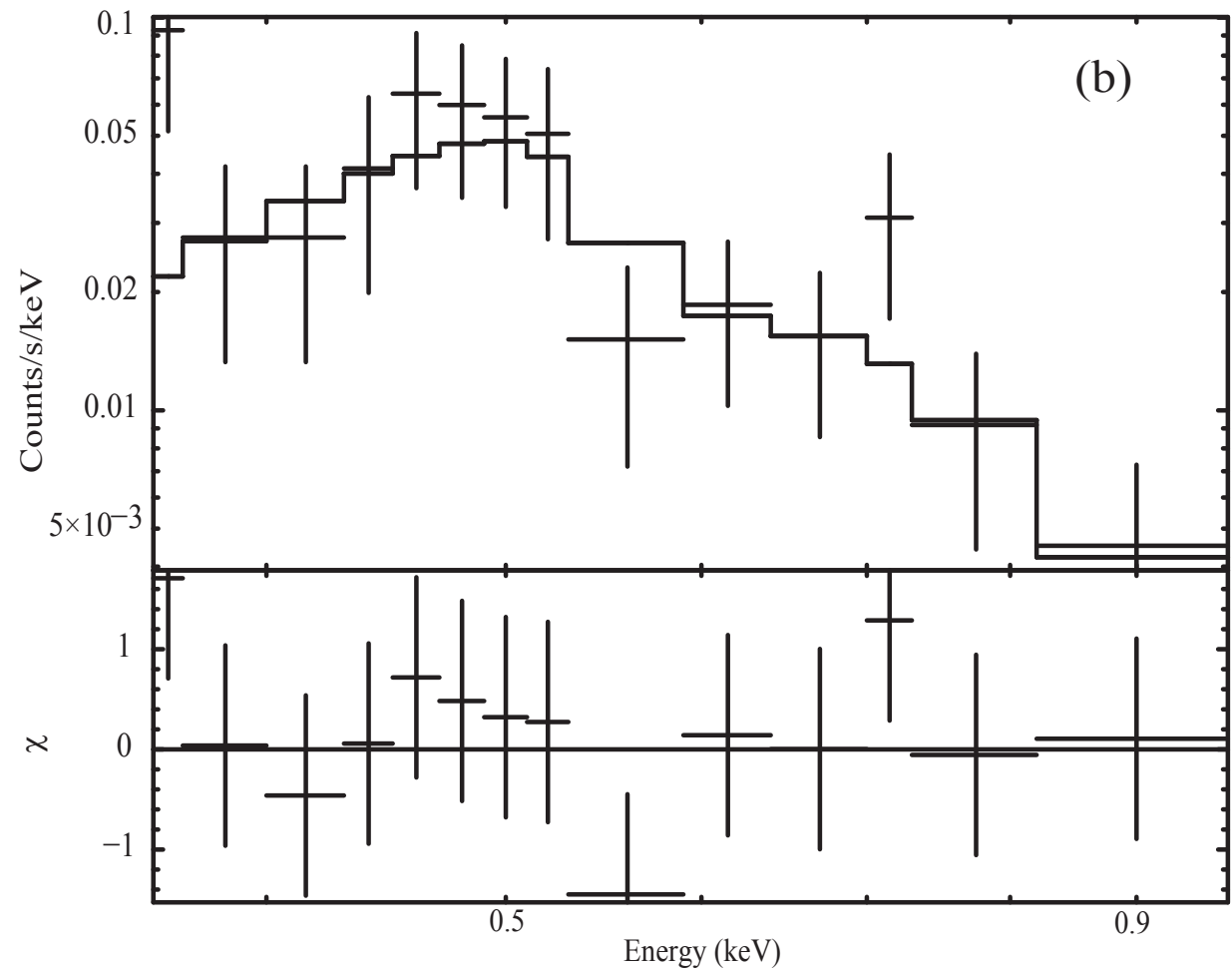

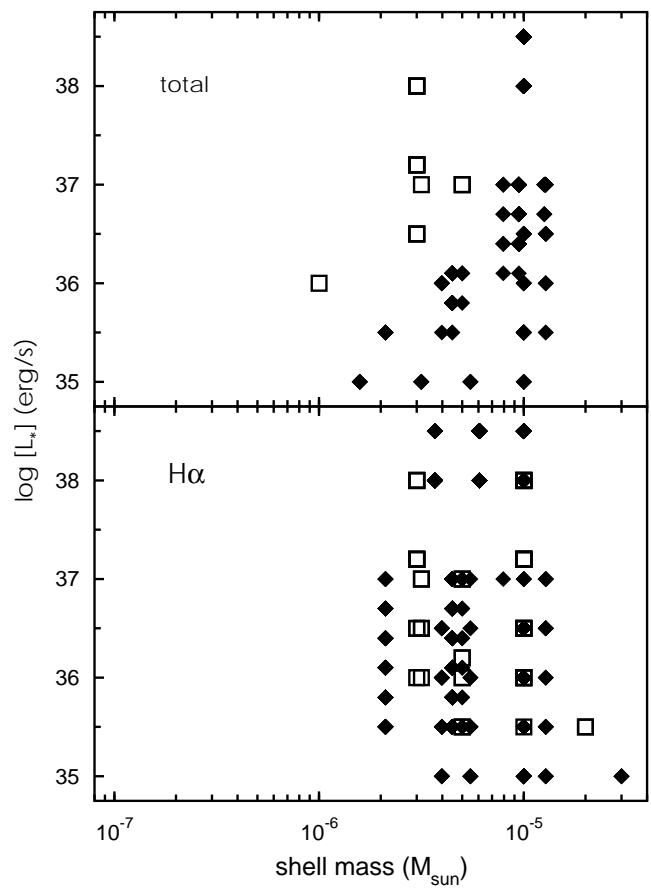